\documentclass[aps,prd,preprint,nofootinbib]{revtex4-1}
\usepackage{amsfonts}
\usepackage{mathrsfs}
\usepackage{graphicx}
\usepackage{amsmath}
\usepackage{amssymb}
\usepackage{epsfig}
\usepackage{graphicx}
\usepackage{slashed}
\usepackage{color}
\usepackage{multirow}
\usepackage{tabularx} 
\usepackage{braket}
\usepackage{bm}
\usepackage{}
\usepackage{caption}
\newcommand{\appropto}{\mathrel{\vcenter{\offinterlineskip\halign{\hfil$##$\cr
				\propto\cr\noalign{\kern2pt}\sim\cr\noalign{\kern-2pt}}}}}

\usepackage{ulem}
\usepackage{adjustbox}

\usepackage{stmaryrd}
\newcommand{\bqa}{\begin{eqnarray}}
	\newcommand{\eqa}{\end{eqnarray}}
\newcommand{\beq}{\begin{equation}}
	\newcommand{\eeq}{\end{equation}}
\allowdisplaybreaks[1]

\graphicspath{{fig/}{dia/}} \DeclareGraphicsExtensions{.eps}
\begin{document} 
\title{
Hidden strangeness in meson weak decays to baryon pair
}
\author{Chao-Qiang Geng$^{1,2}$, Xiang-Nan Jin$^{1,2}$\footnote{
xnjin@ucas.ac.cn	
}, Chia-Wei Liu$^{3}$ and Xiao Yu$^{1,2}$}
\affiliation{$^1$School of Fundamental Physics and Mathematical Sciences, Hangzhou Institute for Advanced Study, UCAS, Hangzhou 310024, China\\
	$^2$University of Chinese Academy of Sciences, 100190 Beijing, China\\
 $^3$Tsung-Dao Lee Institute, Shanghai Jiao Tong University, Shanghai 200240, China}


\date{\today}

\begin{abstract}
Our study focuses on the weak decay of \( D_s^+ \to p \overline{n} \), which is the only possible two-body baryonic decay in the \( D \) meson system. 
An analysis using perturbative quantum chromodynamics (pQCD) is challenging in this decay due to the small amount of energy released. In particular,  
naive factorization, suppressed by the light quark masses,  results in a minor contribution to this channel. In the framework of final state interactions, the hidden strangeness in the intermediate state naturally avoids this chiral suppression from light quark masses. The branching fraction is predicted to be \( {\cal B}(D_s^+ \to p\overline{n}) = (1.43 \pm 0.10) \times 10^{-3} \),  in agreement with the experimental value of \( (1.22 \pm 0.11) \times 10^{-3} \).
We also analyze the decays of \( B \) mesons into two charmed baryons involving annihilation-type topological diagrams. In these decays, we conduct a joint analysis of naive factorization and final state interactions.  Using the experimental upper bound of \( {\cal B}(B_s^0 \to \Lambda_c^+ \overline{\Lambda}_c^-) < 8 \times 10^{-5} \), we set a constraint on the coupling constant \( g_{D^+ \Lambda_c^+ n} < 7.5 \).
Final state interactions lead to a prediction of the decay parameter \( \gamma(B_s^0 \to \Lambda_c^+ \overline{\Lambda}_c^-) > 0.8 \), whereas pQCD predicts it to be negative.
We propose future measurements of \( B^0 \to \Xi_c^+ \overline{\Xi}_c^- \), predicting a significant \( SU(3)_F \) breaking effect with \( \frac{{\cal B}(B^0 \to \Xi_c^+ \overline{\Xi}_c^-)}{{\cal B}(B_s^0 \to \Lambda_c^+ \overline{\Lambda}_c^-)} = 1.4\% \), contrary to the naive estimate of \( 5.3\% \).
We strongly recommend   future measurements.

\end{abstract}

\maketitle

\section{Introduction}

The decays of low-lying \( D \) mesons provide unique opportunities to examine the interplay between perturbative and nonperturbative physics, as well as between weak and strong interactions~\cite{Fajfer:2003ag,Cheng:2024hdo,Petrov:2024ujw, Li:2012cfa}. Notably, \( D_s^+ \to p \overline{n} \) is the only known two-body baryonic decay in the \( D \) meson system~\cite{Pham:1980dc,Bediaga:1991eu,Chen:2008pf,Hsiao:2014zza}.
In 2008, the CLEO collaboration first reported the branching ratio as \( {\cal B}(D_s^+ \to p \overline{n}) = (1.30 \pm 0.36^{+0.12}_{-0.16}) \times 10^{-3} \)~\cite{CLEO:2008aum}, which is a thousand times larger than the theoretical prediction from short-distance (SD) contributions~\cite{Chen:2008pf,Pham:1980dc}. 
In 2019, the BESIII collaboration provided a new measurement, reporting \( {\cal B}(D_s^+ \to p \overline{n}) = (1.21 \pm 0.10 \pm 0.05) \times 10^{-3} \)~\cite{BESIII:2018cfe}, which confirmed the previous experimental data. The significant discrepancy between experimental and theoretical results has sparked considerable interest, leading to further exploration of possible explanations for this anomaly.
For this study, SD refers to the perturbative regime of QCD that treats quarks as the primary entities, with their contributions approximately captured by naive factorization. Conversely, long-distance (LD) corresponds to  physics that treats hadrons as the main entities. Its contributions to amplitudes can be calculated by considering the rescattering of hadrons.

\begin{figure}[htb]
	\begin{center}
		\includegraphics[width=0.3 \linewidth]{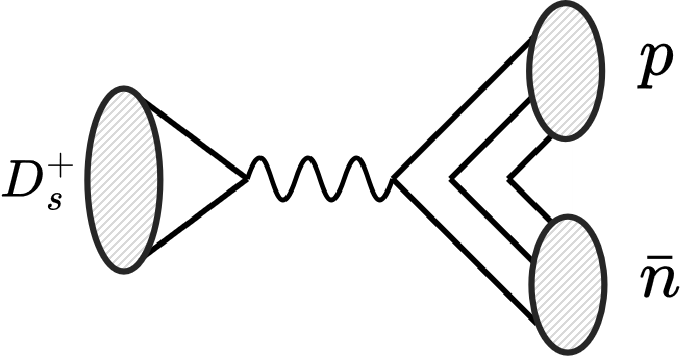}
				\includegraphics[width=0.3 \linewidth]{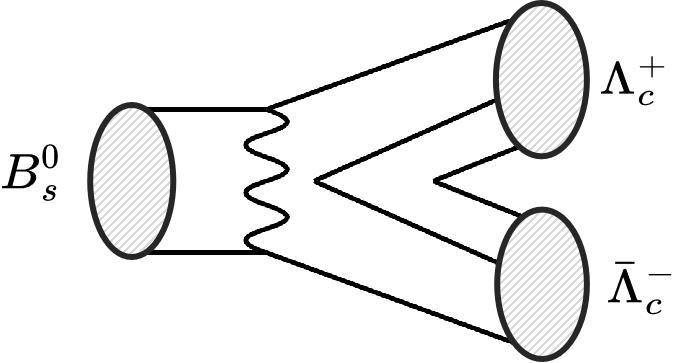}
		\caption{ 
The only possible  quark flow  diagrams for 
			$D_s^+ \to p \overline{n}$ and $B_s^0 \to \Lambda_c^ + \overline{\Lambda}_c^-$.  
		}
		\label{sd}
	\end{center}
\end{figure}

To understand why   $D_s^+ \to p \overline{n}$ is suppressed at  SD, we start with the effective Hamiltonian responsible for the quark-level transition $c \to s \overline{d} u$, which is given by~\cite{Buchalla:1995vs}
\begin{equation}
{\cal H}_{eff} 
= \frac{G_F}{\sqrt{2}}
V_{cs}^* V_{ud}
\left[ 
c_1 (\overline{u} d)_{V-A}
(\overline{s} c)_{V-A}
+ 
c_2 (\overline{s} d)_{V-A}
(\overline{u} c)_{V-A}
\right]\,,
\end{equation}
where $G_F$ is the Fermi constant, $V_{qq'}$ represents the Cabibbo-Kobayashi-Maskawa matrix element, $c_1$ and $c_2$ are the Wilson coefficients, and $(\overline{q_4} q_3)_{V-A} (\overline{q_2} q_1)_{V\pm A}$ denotes $(\overline{q_4} \gamma_\mu (1 - \gamma_5) q_3)(\overline{q_2} \gamma^\mu (1 \pm \gamma_5) q_1)$. 
Under naive factorization, as depicted in Fig.~\ref{sd}, the amplitude is expressed as
\begin{equation}
	{\cal A} _{\text{fac}} (D_s^+ \to p \overline{n})
	= i \frac{G_F}{\sqrt{2}}
	\left(
	c_1 + \frac{1}{N_c}c_2 
	\right)
	V_{cs}^* V_{ud}
	f_D q^\mu 
	\langle
	p \overline{n} | \overline{u} \gamma_\mu (1 - \gamma_5) d | 0 \rangle,
\end{equation}
where $q^\mu$ is the 4-momentum of   $D_s^+$, and $N_c$ is the effective color number. Applying the equation of motion, one finds that the amplitude must be proportional to $m_{u,d}$, which is too small compared to the experimental data. At next-to-leading order in quantum chromodynamics (QCD), two hard gluons must attach to $D_s^+$ and the final state, also resulting in a small branching fraction. 
The core issue is that there is no strange quark in the final state, and the current operator $\overline{s} \gamma^\mu (1 - \gamma_5) c$ must attach to  $D_s^+$, introducing $q^\mu$ and leading to the chiral suppression.
Interestingly, the chiral suppression rule works well in ${\cal B}( B_s^0 \to p\overline{p})$, which is predicted to be less than $10^{-10}$~\cite{Jin:2021onb}. The prediction made in 2021 was soon tested the following year at LHCb, yielding a tiny upper bound of $5.1\times 10^{-9}$~\cite{LHCb:2022oyl}. The reason behind this is that a huge amount of energy is released in the decay of $B_s^0 \to p\overline{p}$, where the degrees of freedom are quarks and gluons. The arguments based on SD are expected to hold. On the other hand, only a tiny amount of energy is released in $D_s^+ \to p \overline{n}$, and the degrees of freedom should be hadrons instead. The arguments based on SD may fail.

Final state interactions (FSIs), are very important at the charm scale~\cite{Qin:2013tje,Qin:2020zlg,Jia:2024pyb,Bediaga:2022sxw,Pich:2023kim}. A typical feature is that quarks may be created and then annihilated before entering the final states~\cite{Fajfer:2003ag}.   Consider the FSI process depicted in Fig.~\ref{fig:two}. The blobs represent the SD transition of $D_s^+ \to P^0 P^+$, induced by ${\cal H}_{eff}$. The quark-level diagrams inside the blob are shown in Fig.~\ref{dk} for $( P^0,P^+) = ( \eta, \pi^+  ), (  \overline{K}^0,K^+ )$. The $s \overline{s}$ pair produced at SD  is hidden in the intermediate hadrons and annihilated before entering the final states~\cite{Fajfer:2003ag}. Clearly, the leading-order FSI amplitudes do not suffer from the chiral suppression. Although the use of FSI to explain the data was proposed years ago~\cite{Chen:2008pf}, an actual calculation had not been carried out until now.

Interestingly, $B_s^0 \to \Lambda_c^+ \overline{\Lambda} _c^-$ exhibits the same feature of small released energy and is solely driven by the $W$-annihilation topology~\cite{Hsiao:2023mud,Cheng:2005vd}, as depicted in Fig.~\ref{sd}. 
However, as we will shortly show, its SD amplitude is not subject to the chiral suppression due to the large mass of the charm quark.
In this work, we study these processes collectively. In Sec. II, we focus on $D_s^+ \to p \overline{n}$. In Sec. III, we examine $B_s^0 \to \Lambda_c^+ \overline{\Lambda} _c^-$ and $B^0 \to \Xi_c^+ \overline{\Xi}_c^-$. Our conclusions are presented in Sec. IV.

\section{$D_s^+  \to p \overline{n}$ }

\begin{figure}[t]
	\begin{center}
		\includegraphics[width=0.35 \linewidth]{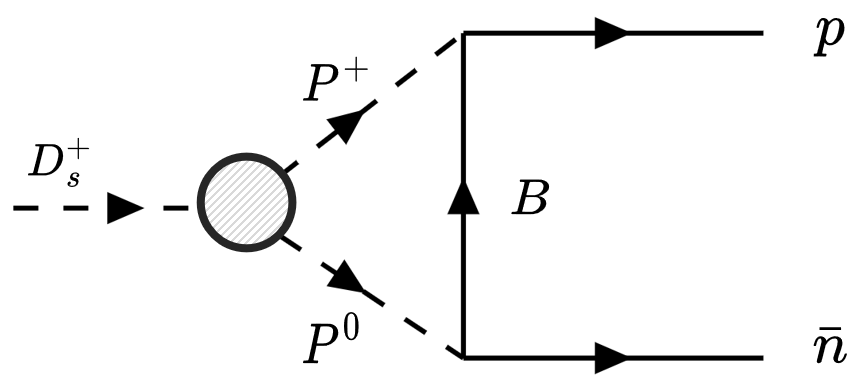}
		\includegraphics[width=0.35 \linewidth]{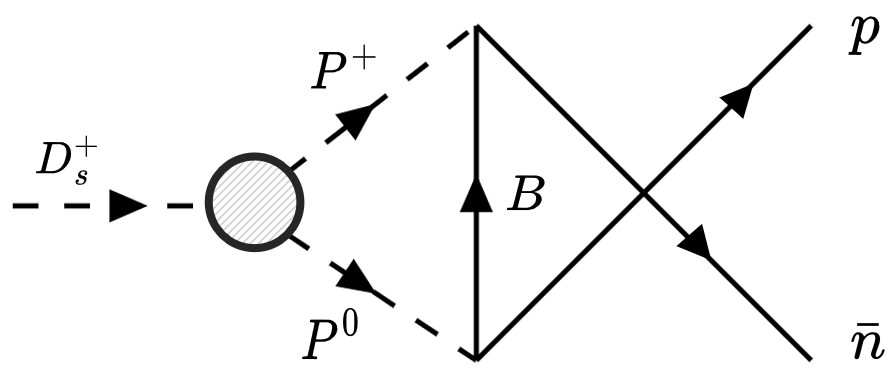}
		\caption{  
	The LD triangular diagrams for 		
	$D_s^+ \to p \overline{n} $. The blob stands for the insertion of ${\cal H}_{eff}$ with the underlying SD transition at   quark level. 
The considered intermediate hadrons are $(P^0,P^+) = (\overline{K}^0,  K^+)$ or 
$(\eta,  \pi^+)$, and $B \in (\Lambda,\Sigma^{0,+},p,n)$. 
		}
		\label{fig:two}
	\end{center}
\end{figure}

\begin{figure}[b]
	\begin{center}
		\includegraphics[width=0.3 \linewidth]{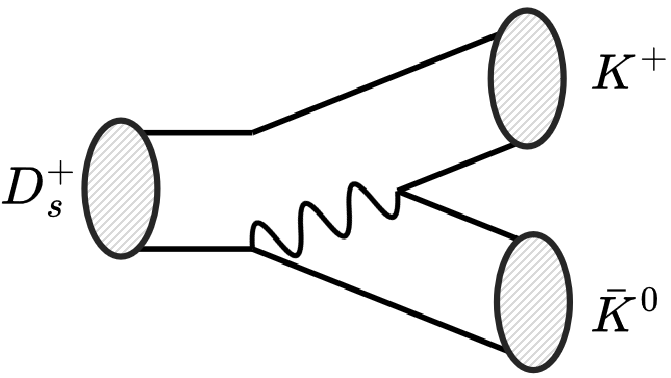}
		\includegraphics[width=0.3  \linewidth]{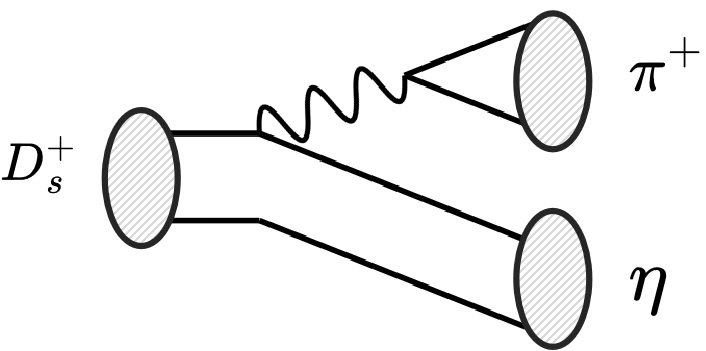}
		\caption{ 
			The SD transitions of 
			$D_s^+ \to K^+ \overline{K}^0$ and $D_s^+ \to \pi^+ \eta$. 
		}
		\label{dk}
	\end{center}
\end{figure}

\begin{figure}[t]
	\begin{center}
		\includegraphics[width=0.32 \linewidth]{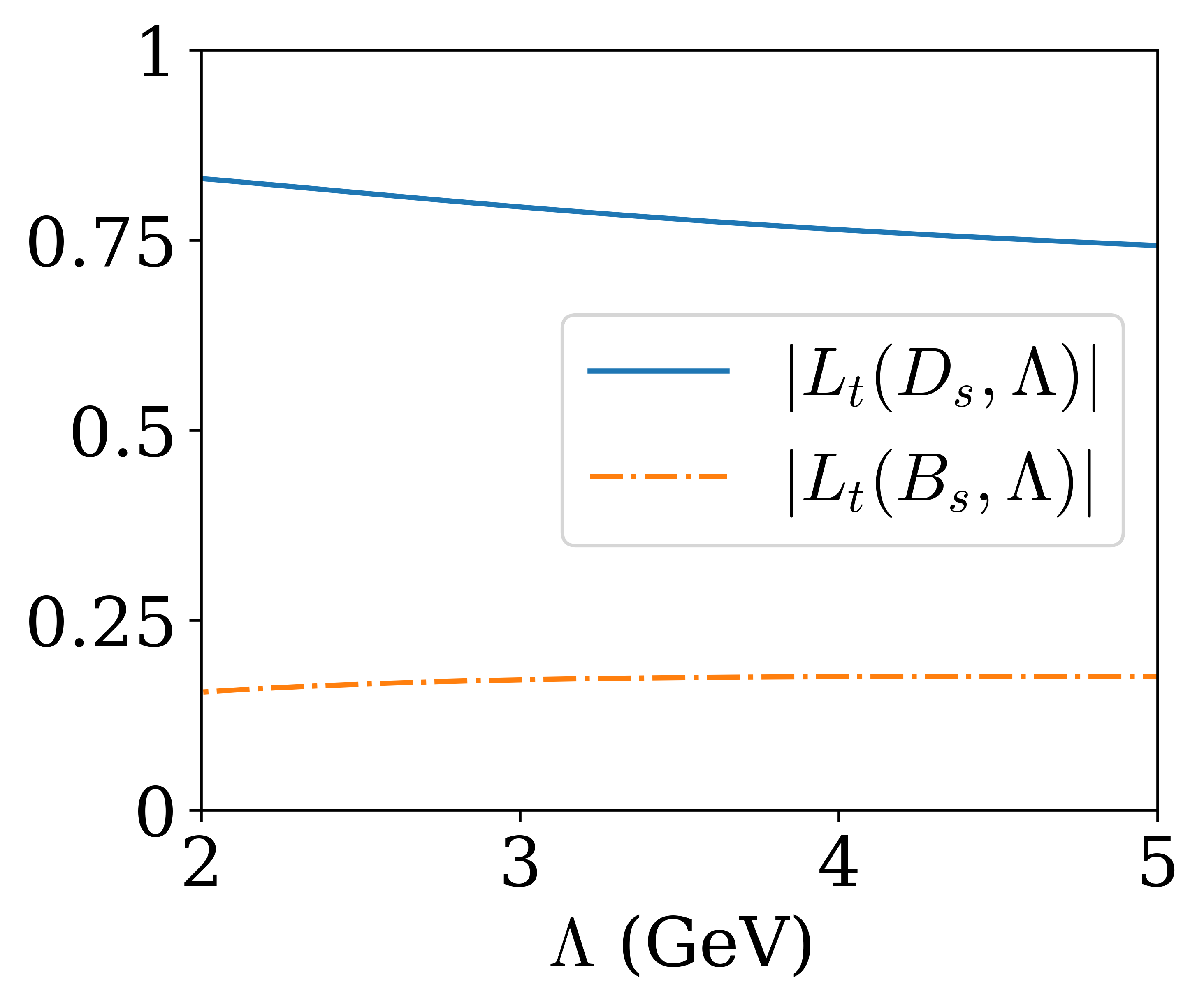}
		\caption{ 
$\Lambda$ dependencies of 
the loop integrals in $D_s^+ \to p \overline{n}$ and $ 
B_s^0 \to \Lambda_c^+ \overline{\Lambda}_c^-$ in the range of 2 GeV$<\Lambda<$5 GeV, denoted as 
$L_t(D_s, \Lambda)$ and 
$L_t(B_s, \Lambda)$, respectively. 
		}
		\label{fig4}
	\end{center}
\end{figure}

The amplitude is in general parametrized by $S$ and $P$ waves:
\begin{eqnarray}
	\langle p\overline{n}| {\cal H}_{eff} | D_s^+ \rangle = i \overline{u}_p (A+B\gamma_5) v_n , 
\end{eqnarray}
with $u_p$ and $v_n$ being the Dirac spinors of $p$ and $\overline{n}$.  
As shown, we consider the SD weak transition  of  $D_s^+ \to \overline{K}^0  K^+ $ and $D_s^+ \to  \eta \pi^+ $, which conserve parity.
Therefore, we can anticipate that  $B=0$ at LD. According to  Fig.~\ref{fig:two}, we have
\begin{eqnarray}
	A &=& \sum_{P^0,P^+,B}T_{ D_s^+   P^0 P^+} ^{\text{weak}} 
	\Big( 
	  g_{P^0  B n } g_{P^+  pB} L_t (m_{P^0},m_B,m_{P^+},p_p,p_n)
\nonumber\\	  && 
	 \qquad\qquad+   
	   g_{P^+  Bn } g_{P^0 pB} L_u (m_{P^0},m_B,m_{P^+},p_p,p_n)\Big), 
\end{eqnarray}
where $(P^0,P^+) = (\overline{K}^0,  K^+)$ or 
$(\eta , \pi^+)$, and $B \in (\Lambda,\Sigma^{0,+},p,n)$. 
The  SD weak transition of $P_0  \to P_1P_2$ is defined by 
\begin{equation}
T^{\text{weak}} _{ 
P_0   P_1 P_2} 
	= \langle P_1 P_2 | {\cal H}_{eff} | P_0 \rangle\,,
\end{equation}
while the hadron couplings are 
\begin{equation}
	{\cal L}_{eff} = \sum_{P,B_1,B_2} 
	- i g_{P B_1B_2} P \left( 
	\overline{B}_1 \gamma_5 B_2\right) + H.c.. 
\end{equation}
The t-channel loop integral  is 
\begin{equation}\small
	L_t(m_1,m_2,m_3, p_1,p_2)
	= 
	\int \frac{d^4q}{(2\pi)^4}
	\frac{
		-(q -p_1)^\mu \gamma_\mu +  m_2
	}{
		(q^2 - m_1^2)
		[(q -p_1)^2 -  m_2^2]
		[(q-p_1-p_2) ^2 -  m_3^2] 
	}\,,
\end{equation}
while the u-channel $
	L_u (m_1,m_2,m_3 , p_1,p_2)
	= 
-	L_t(m_1,-m_2, m_3, p_2,p_1 )
$. 
The analytical form of the loop integral is obtained using Package-X~\cite{Patel:2016fam}.

In practice, FSI integrals often diverge, necessitating the manual inclusion of a cutoff. The UV divergence of the integral arises because the hadron picture is not applicable in the high-energy region, where the theory should  be described by quarks and gluons. To account for the limitations of the hadron description, the common practice in the literature is to include
\begin{equation}\label{Lamb}
	\frac{\Lambda^2}{\Lambda^2 - (q - p_2)^2},
\end{equation}
inside the FSI integral.
The parameter \( \Lambda \) must be large enough to not affect  the low-energy physics of interest but small enough to exclude contributions from high energies. These conflicting requirements often undermine the predictive power of the FSI framework. For reliable results, the numerical outcomes should depend minimally on the specific value of \( \Lambda \).

We remark that our FSI integral of \( L_t \) converges  even without a cutoff, indicating that it receives little contribution from the high-energy region. Nevertheless, it is still interesting to examine the impact  of a  cutoff.
Setting the other parameters to their physical values and including Eq.~\eqref{Lamb} into the integrand, we plot the dependence of \( \Lambda \) in Fig.~\ref{fig4}, denoted as \( |L_t(D_s,\Lambda)| \) for \( 2\,\text{GeV} < \Lambda < 5\,\text{GeV} \). We have also included the integrals encountered in \( B_s^0 \to \Lambda_c^+ \overline{\Lambda}_c^- \), denoted as \( |L_t(B_s,\Lambda)| \). Both vary by less than \( 10\% \), and we conclude that the numerical results are stable. In the following, we take \( \Lambda \to \infty \).

In this work, 
we determine 
$T^{\text{weak}} _{ 
	P_0   P_1 P_2} $ using the experimental data with 
\begin{equation}
\Gamma (P_0 \to P_1 P_2) = \frac{|\vec{p}|}{8 \pi m_{0}^2} |T^{\text{weak}} _{ 
	P_0   P_1 P_2}  |^2\,,
\end{equation}
where $m_0$ is the mass of $P_0$. 
From ${\cal B}(D_s^+ \to \overline{K}^0 K^+) = (2.94 \pm 0.06)\%$ and 
${\cal B}(D_s^+ \to \eta \pi^+) = (1.67 \pm 0.09)\%$~\cite{ParticleDataGroup:2024cfk}, we derive  
\begin{equation}
( 	T^{\text{weak}} _{ 
D_s^+   \overline{K}^0 K^+},
	T^{\text{weak}} _{ 
	D_s^+ \eta \pi^+ } ) = (0.180 \pm 0.02, -0.132 \pm 0.003) G_F \text{GeV}^3\,.
\end{equation}
They are chosen to be opposite in sign in accordance with naive factorization. 
The adopted hadron couplings are
\begin{eqnarray}
	&& g_{\overline{K}^0   \Sigma^0 n } = g_{ K^+ p {\Sigma}^0} 
	= \frac{1}{\sqrt{2}}g_{ K^+ \Sigma ^+ n } 
	= -\frac{1}{\sqrt{2}} g_{ \overline{K} ^0  \Sigma ^+  p  } 
	= 3.215\pm 0.163 \,,  \nonumber \\
	&&
	g_{  \overline{K}^0  \Lambda  n  }  = - g_{K^+ p {\Lambda}} =  9.228 \pm 0.209\,,~~~~\frac{1}{\sqrt{2}} g_{\pi^+  pn } =   12.897\pm0.047  \,.
\end{eqnarray}
The  above couplings   are extracted from the Goldberger-Treiman relations~\cite{General:2003sm}. For $g_{\eta NN}$, the central value
 is adopted from the photoproduction~\cite{JPAC:2016lnm},
 given as $g_{\eta  {n}n} = g_{\eta  p {p}} =  0.89 \pm 0.18 $
  with $20\%$ uncertainties   included to be conservative. 

Recalling  the decay branching width  is given by  
\begin{equation}   \small 
	\Gamma ( D_s^+ \to p \overline{n} ) = \frac{|\vec{p}| }{4 \pi m_{D_s^+}^2 }  
	\left\{
	\left[
	m_{D_s^+}^2 - (m_p+m_n)^2 
	\right]
	|A |^2
	+
	\left[  m_{D_s^+}^2 - (m_p- m_n)^2 
	\right]
	|B |^2
	\right\}\,,
\end{equation}
and plugging everything, 
we find 
\begin{equation}
	{\cal B}(D_s^+ \to p \overline{n} ) = ( 1.43\pm 0.10)  \times 10 ^{-3} \,,
\end{equation}
which is well consistent with the data of $(1.21\pm 0.10\pm0.05)\times 10^{-3}$. 
 Naively, the kinetic factor in front of   
 $|B|^2$ is ten times larger than the one of  $|A|^2$ and one may expect that the $P$ wave dominates the decays. 
 Nevertheless, our result shows that the $S$ wave induced by the FSIs alone saturates the experimental large branching fractions.
 In Table~\ref{final }, we also include the 
 SD 
 prediction of ${\cal B}_P$ from Ref.~\cite{Chen:2008pf}.  
 Here, ${\cal B}_{S(P)}$ stands for the branching fractions stemmed in  $S(P)$ waves. 
 It is interesting to point out that the SD and LD contribute separately to 
 $S$ and $P$ waves.

\begin{table}[t]  
	\setlength\tabcolsep{4pt}
	\caption{Collections of 
		the branching fractions 
		 compared to the data~\cite{ParticleDataGroup:2024cfk}.
		 The prediction of ${\cal B}_P(D_s^+ \to p \overline{n} )$ is taken from Ref.~\cite{Chen:2008pf}. 
		   Here, ${\cal B}_{S,P}$ stand  for the branching $S$ and $P$  wave parts of the branching fractions, respectively. }
	\label{final }
	\resizebox{1\textwidth}{!}{
	\begin{tabular}{l|cccc}
		\hline
		Channel  & ${\cal B}_{S} $
		&  ${\cal B}_{P}$& ${\cal B}_{\text{tot}}$
		& ${\cal B}_{\text{tot}}^{\text{exp}} $   \\
		\hline
		$D_s^+ \to p \overline{n} $& $( 1.43 \pm 0.10   ) \times   10 ^{-3} $ 
		& $( 0.4 ^{+1.1}_{-0.3} )\times  10 ^{-6}  
		$ & $( 1.43 \pm 0.10 )   \times 10 ^{-3} $  & $(1.22\pm 0.11  ) \times   10 ^{-3} $
		\\ 
		\hline  
		$ B_s^0 \to \Lambda_c^+ \overline{\Lambda} _c^- $
		&$>  4.2   \times 10 ^{-5 }$  
		& $(3.9 \pm 1.5 )\times   10^{-6}$& $>  4.7 \times   10^{-5}$& $<8 \times 10 ^{-5}$
		\\ 
		$ B_s^0 \to \Sigma _c^+ \overline{\Sigma} _c^- $
		&$ < 10 ^{-6} $   &$(3.4\pm 1.6) \times 10^{-6}$ &$(3.9\pm 2.1)  \times  10^{-6}$& $-$
		\\ 
		$ B^0 \to \Xi _c^+ \overline{\Xi} _c^- $
		& $>   7  \times 10 ^{-5 } $
		&$ (1.3 \pm 0.5 ) \times 10 ^{-7}$  & $>   7  \times 10 ^{-5 } $& $-$
		\\ 
		$ B ^0 \to \Xi  _c^{\prime +}  \overline{\Xi } _c^{\prime  - }  $
		&$<3 \times 10 ^{-8}$ &$(8 \pm 3 )\times 10^{-8}$ &$( 1.0 \pm 0.4) \times 10^{-7}$& $-$
		\\ 
		\hline
		\hline
	\end{tabular}
}\end{table}

\section{$B_s^0 \to \Lambda_c^+ \overline{\Lambda}_c^-, \Sigma_c^+ \overline{\Sigma}_c^-$ and 
$B ^0 \to \Xi_c^+ \overline{\Xi}_c^-
,\Xi_c^{\prime +} \overline{\Xi}_c^{\prime -}$
}

By the same token, we consider the re-scattering processes of $B_{s} \to Y_c^+ \overline{Y}_c^{  -}$ with 
$B_s \in \{
B_s^0 , \overline{B}_s^0 , B_{sL}^0 
, B_{sH}^0 
\}$ and 
$Y_c^+   \in  \{\Lambda _c^+ ,   \Sigma_c^+       \} $.
The amplitudes in general read as 
\begin{equation}\label{zhou}
{\cal M} ( B_s^0 \to Y_c ^+ \overline{Y}_c^{  -} ) 
= i \overline{u}_c \left(
A_{LD} + B_{SD }  \gamma_5
\right) v _ c \,. 
\end{equation}
The subscripts are due to the anticipation that \( S \) and \( P \) waves mainly contribute through LD and SD, respectively. This feature naturally avoids the double-counting problem in the joint consideration of SD and LD, since they contribute to different partial waves and do not interfere.

Notably, $A_{LD}/B_{SD}$ would be $CP$-violating for $B_{s} = B_{sH}^0/B_{sL}^0$.  
The tininess of $CP$ violation in the charmful $b \to s$ transitions leads to a selection rule in the partial waves
\begin{equation}\small 
	{\cal B}(B_{sH}^0
	\to Y_c^+ \overline{Y}_c^- ) 
	= 
	2 {\cal B}_P(B_{s}^0
	\to Y_c^+ \overline{Y}_c^- )\,,~~~~
	{\cal B}(B_{sL}^0
	\to Y_c^+ \overline{Y}_c^- ) 
	= 
	2 {\cal B}_S(B_{s}^0
	\to Y_c^+ \overline{Y}_c^- )\,. 
\end{equation}
It allows us to disentangle the $S$-wave and $P$-wave contributions.

The effective Hamiltonian governing the charmful $b\to s$ transitions is given by~\cite{Buchalla:1995vs} 
\begin{eqnarray}
	{\cal H}_{eff} 
	&=& \frac{G_F}{\sqrt{2}}
\Bigg\{ 
	V_{cb} V_{cs}^*
\left[ 
	c_1 (\overline{s}  c )_{V-A}
	(\overline{c}  b)_{V-A}
	+ 
	c_2 (\overline{c}  c )_{V-A}
	(\overline{s}  b)_{V-A}
	\right]  \nonumber \\
	&+&
V_{tb} V_{ts}^*
\Big[ 
c_3 (\overline{s}  b )_{V-A}
(\overline{c}  c)_{V-A}
+ 
c_4 (\overline{s}_\alpha  b_\beta  
)_{V-A}
(\overline{c}_\beta  c_\alpha)_{V - A}
\\ &&\qquad + 
c_5 (\overline{s} b  )_{V-A}
(\overline{c} c)_{V+A}
+ 
c_6 (\overline{s}_\alpha  b_\beta  
)_{V-A}
(\overline{c}_\beta  c_\alpha)_{V+ A}
\Big] 
\Bigg\} \,.\nonumber 
\end{eqnarray}
Here, $\alpha$ and $\beta$ are the color indices. In this work, we omit the color indices when the quark and anti-quark of the current operators share the same color, {\it i.e.},
$
(\overline{q}' q)_{V\pm A} = (\overline{q}_\alpha' q_\alpha)_{V\pm A}.$
As   again, the final state is strangeless, there is only one 
$W$-annihilation
 topological diagram.
A naive factorization from the SD
for $B_s = B_s^0$ 
would give us 
\begin{equation}
B_{SD }
	=  \frac{G_F}{\sqrt{2}}
	\left(
	c_2 + \frac{1}{N_c}c_1 
	\right)
	V_{cb} V_{cs}^*
	f_{Bs}  2 m _c  
f_s (m_{B_s}^2 ) .
\end{equation}
with the form factor defined as $
\langle Y_c^+ \overline{Y}_c^{ - }  | \overline{c} \gamma_5 c | 0 \rangle = 
f_s(q^2) \overline{u}_c \gamma_5 v_c $. 
The $c_{3\sim 6}$ terms  are omitted for
the large cancellations
 $c_{3,4}+c_{5,6} \approx 0 $. It is often found in the literature that naive factorization is not trustworthy for color-suppressed diagrams. A common practice is to treat $N_c$ as an effective free parameter, at the cost of predictive power. Aiming to show that the SD contributions are negligible, we take a large value of $a_2 = (c_2 + c_1/N_c) = 1$. On the other hand, the form factor is estimated using the homogeneous bag model combined with the $3P^0$ pair creation model. It varies little across decays, and we find $f_s(m_B^2) \approx (1.5 \pm 0.3)\%$.
Calculation details and model parameter inputs can be found in Ref.~\cite{Geng:2023yqo}. As emphasized in our framework, the $S(P)$-wave branching fractions are exclusively contributed by the LD(SD), and one can directly read off the contributions of SD and LD from ${\cal B}_{S}$ and ${\cal B}_P$, respectively in Table~\ref{final }.

\begin{figure}[t]
	\begin{center}
	\includegraphics[width=0.32 \linewidth]{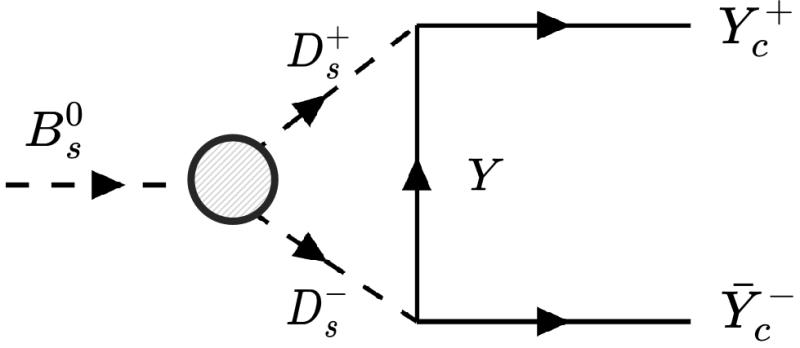}
		\caption{ 
	The LD triangular diagram for 		
$B_s^0 \to Y_c^+ \overline{Y}_c^- $ with $Y_c^+= \Lambda _c^+ ,\Sigma _c^+$
for  $Y = \Lambda , \Sigma$, respectively.
	}
		\label{fig2}
	\end{center}
\end{figure}

The FSI  
diagram 
of $B_s^0 \to D_s^+  {D}_s^- \to 
Y_c^+  \overline{Y}_c^-$  is depicted in Fig.~\ref{fig2}. The amplitudes 
are given by 
\begin{equation}   
 A _{LD} 
=
T^{\text{weak}} _{ 
B_s^0   D_s^+  {D}_s^ -}    (g_{D_s^+  Y_c^+  Y })^2  L_t (m_{D_s },m_Y ,m_{ D_s  },p_{Y_c^+},p_{ \overline{Y}  _c^-})\,. 
\end{equation}
The experimental input of 
$
 {\cal B}(B_s^0 \to D_s^+ {D} _s^-) =  (4.4\pm 0.5)  \times 10 ^{-3} 
$~\cite{ParticleDataGroup:2024cfk}, corresponds to $
T^{\text{weak}} _{ 
	B_s^0   D_s^+   {D}_s^ -} = 
( 7.5 \pm 0.5)  \times 10 ^{-2}  G_F \text{GeV}^3 \,. 
$
For  $g_{D_s ^+ Y_cY}$,  we use the $SU(3)_F$ relation of 
$
g_{D_s ^+   \Lambda^+  _c\Lambda }
= 
\sqrt{
2/3} 
g_{D^+   \Lambda_c^+ n }
  $ and $ 
g_{D_s ^+  \Sigma ^+  _c\Sigma^0  }
= 
\sqrt{   2 } 
g_{D^+  \Sigma_c ^+ n }\,,
$
and arrive at 
\begin{eqnarray}
{\cal B}_S(B_s^0 \to \Lambda_c^+ \overline{\Lambda}_c^-)
&=& 
(g_{D^+ \Lambda^+_cn})^4
\times \left( 2.53\pm0.29 \right)   
\times 10 ^{-8}\,,  \nonumber\\
{\cal B}_S(B_s^0 \to \Sigma_c^+ \overline{\Sigma}_c^-)
&=& 
(g_{D^+ \Sigma ^+_cn})^4
\times \left( 2.38 \pm 0.27\right) 
\times 10 ^{-7}\,.
\end{eqnarray}
Using the QCD light-cone sum rule inputs
  $
g_{D^+   \Lambda_c^+ n } =  10.7^{+5.3}_{-4.3} $ and 
$g_{D^+ \Sigma_c ^+ n } = 1.3 ^{+1.0}_{-0.9} $~\cite{Khodjamirian:2011sp}, 
the net results  are 
summarized in Table~\ref{final }.  
Only lower and upper  bounds are obtained in $B_s^0 \to \Lambda_c^+ \overline{\Lambda}_c^-$  and $B_s^0 \to \Sigma_c^+ \overline{\Sigma}_c^-$, respectively, due to 
the hadron coupling   is   poorly determined.
Alternatively,
 the experimental upper bound of $ 
 {\cal B}(B_s^0 \to \Lambda_c ^+ \overline{\Lambda}_c^- ) < 
 8 \times 10 ^{-5} $ 
 set the constraint of 
    $
g_{D^+   \Lambda_c^+ n } < 7.5$.

Recently, an interesting paper also calculated the process of \( B_s^0 \to \Lambda_c^+ \overline{\Lambda}_c^- \)~\cite{Rui:2024xgc}. The authors found that the experimental upper bound of 
${\cal B}$ 
can be explained by pQCD, with \( \gamma = \left( {\cal B}_S - {\cal B}_P \right) / \left( {\cal B}_S + {\cal B}_P \right) = -0.82^{+0.05}_{-0.03} \) with minimal uncertainties. This result, as anticipated in Eq.~\eqref{zhou}, suggests that the \( P \)-wave is dominated by SD physics. However, by considering the LD contribution, we obtain a significantly different result, with \( \gamma > 0.8 \). The parameter \( \gamma \) can be measured through the cascade decays~\cite{Geng:2023nia}, and such a measurement would shed light on the applicability of pQCD in this region and the contributions of FSIs. We strongly recommend an experimental measurement.

 The same analysis can be performed for $B^0 \to D^+ {D}^- \to X_c^+ \overline{X}_c^-$, with $X_c^+ = \Xi_c^+, \Xi_c^{\prime +}$, by substituting the $s$-quark in the former case with the $d$-quark. Using ${\cal B}( B^0 \to D^+ {D}^-) = (2.11 \pm 0.18) \times 10^{-4}$ as input~\cite{ParticleDataGroup:2024cfk}, the results are also documented in Table~\ref{final }. 
 To examine the $SU(3)_F$ breaking, we define the ratios:
 \begin{equation}
 {\cal R}_M = \frac{\Gamma( B^0 \to D^+ {D}^-)}{\Gamma( B_s^0 \to D_s^+ {D}_s^-)} \,, ~~~
{\cal R}_B = \frac{\Gamma( B^0 \to \Xi_c^+ \overline{\Xi}_c^-)}{\Gamma( B_s^0 \to \Lambda_c^+ \overline{\Lambda}_c^-)} \,, ~~~
{\cal R}_{B'} = \frac{\Gamma( B^0 \to \Xi_c^{\prime +}  \overline{\Xi}_c^{\prime -})}{\Gamma( B_s^0 \to \Sigma_c^+ \overline{\Sigma}_c^-)} \,.
 \end{equation}
 In the limit of U-spin symmetry, we have
 \begin{equation}
 {\cal R}_M^{\text{U}} = {\cal R}_B^{\text{U}} = {\cal R}_{B'}^{\text{U}} = \left| \frac{V_{cd}^2}{V_{cs}^2} \right| \approx 5.3\% \,.
 \end{equation}
 The prediction of ${\cal R}_M^{\text{U}}$ is consistent with the experimental value of ${\cal R}_M^{\mathrm{exp}} = (4.8 \pm 0.7)\%$. However, experimental measurements for baryonic decays are still missing. We predict
 \begin{equation}
 {\cal R}^{\text{FSI}}_B = 1.4\% \,, ~~~ {\cal R}^{\text{FSI}}_{B'} = 2.8\% \,.
 \end{equation}
 Uncertainties in the hadron couplings are canceled in the ratio, allowing for precise determination. The large U-spin breaking in ${\cal R}_B^{\text{FSI}}$ is traced back to the fact that the released energy is of the same order as the strange quark mass. In particular, the decay widths are given by
\begin{equation}
 \Gamma ( B \to {\bf B} \overline{{\bf B}}) = \frac{\left( m_{B }^2 - 4m_{{\bf B}}^2 \right)^{3/2}}{8 \pi m_{B }^2} |A_{LD}|^2 + \frac{\sqrt{ m_{B }^2 - 4m_{{\bf B}}^2 }}{8 \pi } |B_{SD}|^2 \,,
\end{equation}
 where $m_{B_s^0} - 2 m_{\Lambda_c} = 0.76$ GeV and $m_{B^0} - 2 m_{\Xi_c} = 0.34$ GeV. 
  Since the breaking effects in the $S$  and $P$ waves manifest very differently, we conclude that the results from a purely $SU(3)_F$ analysis may be questionable without considering partial waves. Future experiments on $ B^0 \to \Xi_c^+ \overline{\Xi}_c^- $ would be helpful in clarifying this issue.

\section{
Conclusion
}
We have studied the weak decays of a heavy meson into a pair of baryons, focusing on decay channels dominated by annihilation-type diagrams and small released energies. The presence of hidden strangeness in the intermediate state naturally avoids chiral suppression for $D_s^+ \to p \overline{n}$. A cutoff is not needed in our framework,  enhancing the reliability of our predictions.
For $D_s^+ \to p \overline{n}$, the SD contributions are highly suppressed. Using LD triangular FSI, we have predicted ${\cal B}(D_s^+ \to p \overline{n}) = (1.43 \pm 0.10) \times 10^{-3}$, which is consistent with the current data of $(1.22 \pm 0.11) \times 10^{-3}$. For $B$ meson decays, we have found ${\cal B}(B_s^0 \to \Lambda_c^+ \overline{\Lambda}_c^-) > 4.7 \times 10^{-5}$, which is compatible with the experimental lower bound of $8 \times 10^{-5}$. The large uncertainties are due to the poorly known couplings between $D$ mesons and nucleons in the literature. Working backward from the experimental upper bound of $B_s^0 \to \Lambda_c^+ \overline{\Lambda}_c^-$, we derive $g_{D^+ \Lambda_c^+ n} < 7.5$.
We have found that $CP$ symmetry leads to a selection rule such that $B_{sH/L}^0 \to \Lambda_c^+ \overline{\Lambda}_c^-$ is dominated by $P/S$ wave. This allows us to distinguish the SD and LD contributions and provides valuable tests in nonperturbative physics.
	It has been shown that 
	FSIs lead to \( \gamma(B_s^0 \to \Lambda_c^+ \overline{\Lambda}_c^-) > 0.8 \), whereas pQCD predicts it to be negative.
Finally, we recommend that future experiments measure \( B^0 \to \Xi_c^+ \overline{\Xi}_c^- \) to test the FSI mechanism, where a large \( SU(3)_F \) breaking of \( {\cal R}^{\text{FSI}}_B = 1.4\% \) is predicted, contrary to the naive prediction of \( {\cal R}_B^{\text{FSI}} = 5.3\% \). Future measurements would be most welcome.

\begin{acknowledgements}
This work is supported in part by the National Key Research and Development Program of China under Grant No. 2020YFC2201501 and  the National Natural Science Foundation of China (NSFC) under Grant No. 12347103 and No. 12205063.
\end{acknowledgements}

\end{document}